\documentclass[journal]{IEEEtran}
\usepackage[latin9]{inputenc}
\usepackage{array}
\usepackage{url}
\usepackage{amsmath}
\usepackage{amssymb}
\usepackage{graphicx}
\PassOptionsToPackage{normalem}{ulem}
\usepackage{ulem}

\makeatletter

\providecommand{\tabularnewline}{\\}

\usepackage{amsfonts}
\usepackage{multirow}
\providecommand{\U}[1]{\protect \rule{.1in}{.1in}}
\makeatother

\begin{document}

\title{An Analytical Design Optimization Method for Electric Propulsion
Systems of Multicopter UAVs with Desired Hovering Endurance}

\author{Xunhua Dai, Quan Quan, Jinrui Ren and Kai-Yuan Cai \thanks{The authors are with School of Automation Science and Electrical Engineering,
Beihang University, Beijing 100191, China.}}
\maketitle
\begin{abstract}
Multicopters are becoming increasingly important in both civil and
military fields. Currently, most multicopter propulsion systems are
designed by experience and trial-and-error experiments, which are
costly and ineffective. This paper proposes a simple and practical
method to help designers find the optimal propulsion system according
to the given design requirements. First, the modeling methods for
four basic components of the propulsion system including propellers,
motors, electric speed controls, and batteries are studied respectively.
Secondly, the whole optimization design problem is simplified and
decoupled into several sub-problems. By solving these sub-problems,
the optimal parameters of each component can be obtained respectively.
Finally, based on the obtained optimal component parameters, the optimal
product of each component can be quickly located and determined from
the corresponding database. Experiments and statistical analyses demonstrate
the effectiveness of the proposed method. The proposed method is fast
and practical that it has been successfully applied to a web server
to provide online optimization design service for users. 
\end{abstract}


\section{Introduction}

During recent years, multicopter Unmanned Aerial Vehicles (UAVs) are
becoming increasingly popular in both civil and military fields \cite{doyle2013avian,tauro2015large}
including aerial photography, plant protection, package delivery and
other fields. Limited by the battery technology, the flight time (hovering
endurance) of multicopters is still too short for most applications.
Since the performance and efficiency of a multicopter directly depend
on the propulsion system, the design optimization for multicopter
propulsion systems is urgently needed to increase the flight time.

The design optimization problem studied in this paper is to find the
optimal combination of propulsion system components to satisfy the
given hovering endurance requirement, and the obtained propulsion
system should have smaller weight and higher efficiency as possible.
A typical multicopter propulsion system usually consists of four basic
components including the propeller, BrushLess Direct-Current (BLDC)
motor, Electronic Speed Control (ESC) and Lithium Polymer (LiPo) battery
\cite{quan2017introduction}. Traditional methods to determine a propulsion
system are usually based on the experience and trial-and-error experiments.
Considering that there are thousands of component products on the
market, it is a costly and time-consuming work for the traditional
design methods. Meanwhile, in the whole design process of a multicopter
system, the propulsion system need to be repeatedly modified according
to the actual controlled system until all the performance requirements
and safety requirements are satisfied. According to \cite{oktay2013simultaneous,oktay2016simultaneous},
a more efficient way is to simultaneously design the body system (including
the propulsion system) and the control system subject to the optimal
objective and additional constraints. Therefore, the automatic design
and optimization technologies for propulsion systems will be beneficial
for reducing the prototyping needs for the whole multicopter system,
and minimizing development and manufacturing cost. For such reasons,
this paper proposes a simple, practical and automatic design optimization
method to help designers quickly find the optimal propulsion system
according to the given design requirements. 

In our previous work \cite{Shi2017}, based on the mathematical modeling
methods for the components of propulsion systems, a practical method
is proposed to estimate the flight performance of multicopters according
to the given propulsion system parameters. In fact, the study in this
paper is the reverse process of our previous work, namely estimating
the optimal propulsion system parameters according to the given design
requirements. This problem is more complicated and difficult because
the number of design requirements is much less than the number (more
than 15) of the propulsion system parameters.

There are many studies on the mathematical modeling \cite{Harrington2011,Ramana2013,Mccrink2015},
the efficiency analysis \cite{Lawrence2005,Stepaniak2009}, and the
performance estimation \cite{Shi2017,Bershadsky2016a} of multicopter
propulsion systems. To our best knowledge, there are few studies on
the design optimization of multicopter propulsion systems. Most of
them adopt numerical methods (fixed-wing aircraft \cite{Lundstrom2009}
and multicopters \cite{Bershadsky2016a}) to search and traverse all
the possible propulsion system combinations in the database based
on the proposed cost functions. In \cite{Magnussen2014,Magnussen2015},
the multicopter optimization problem is described as a mixed integer
linear program, and solved with the Cplex optimizer. However, these
numerical methods have following problems: i) a large and well-covered
product database is required for a better optimization effect; ii)
the calculation speed is slow when there are large numbers of products
in the database because the amount of product combinations is huge
(the algorithm complexity is $O(n^{4})$, where $n$ is the number
of the database products).

In order to solve the above problems, this paper proposes an analytical
method to estimate the optimal parameters of the propulsion system
components. First, the modeling methods for each component of the
propulsion system are studied respectively to describe the problem
with mathematical expressions. Secondly, the whole problem is simplified
and decoupled into several small problems. By solving these sub-problems,
the optimal parameters of each component can be obtained respectively.
Finally, based on the obtained parameters, selection algorithms are
proposed to determine the optimal combination of the propeller, the
motor, the ESC and the battery products from their corresponding databases.

The contributions of this paper are as follows: i) an analytical method
to solve the design optimization problem of multicopter propulsion
systems is proposed for the first time; ii) the conclusion obtained
through the theoretical analysis has a guiding significance for the
multicopter design; iii) compared to the numerical traversal methods,
the proposed method reduces the algorithm complexity from $O(n^{4})$
to $O(n)$, which is faster and more efficient for practical applications.

The paper is organized as follows. \textit{Section\,\ref{Sec-2}}
gives a comprehensive analysis of the design optimization problem
to divide it into twelve sub-problems. In \textit{Section\,\ref{Sec-3}},
the modeling methods for each component of the propulsion system are
studied the describe the sub-problems with mathematical expressions.
In \textit{Section\,\ref{Sec-4}}, the sub-problems are solved respectively
to obtain the optimal components of the desired propulsion system.
In \textit{Section\,\ref{Sec-6}}, statistical analyses and experiments
are performed to verify the proposed method. In the end, \textit{Section\,\ref{Sec-7}}
presents the conclusions.

\section{Problem Formulation}

\label{Sec-2}

\subsection{Design Requirements of Propulsion Systems}

\label{Sec2-Des}

The role of a propulsion system is to continuously generate the desired
thrust within the desired time of endurance for a multicopter. The
design requirements for a propulsion system are usually described
by the following parameters: i) the number of the propulsion units
$n_{\text{p}}$; ii) the hovering thrust of a single propeller $T_{\text{hover}}$
(unit: N) under the hovering mode when the multicopter stays fixed
in the air; iii) the maximum thrust of a single propeller $T_{\text{max}}$
(unit: N) under the full-throttle mode when the autopilot gives the
maximum throttle signal; iv) the nominal flight altitude $h_{\text{hover}}$
(unit: m); iv) the flight time $t_{\text{hover}}$ (unit: min) under
the hovering mode.

This paper only focuses on studying the design optimization of propulsion
systems with assuming $T_{\text{hover}}$ and $T_{\text{max}}$ are
known parameters. Although the propulsion system parameters $T_{\text{hover}}$
and $T_{\text{max}}$ are usually not directly available, according
to our previous research \cite{Shi2017}, $T_{\text{hover}},T_{\text{max}}$
can be obtained by giving the aerodynamic coefficients, airframe parameters
and the kinematic performance requirements. For example, for common
multicopters, the hovering thrust $T_{\text{hover}}$ can be obtained
by the total weight of the multicopter $G_{\text{total}}$ (unit:
N) as
\begin{equation}
T_{\text{hover}}=\frac{G_{\text{total}}}{n_{\text{p}}}.\label{eq:T0}
\end{equation}
The kinematic performance of a multicopter is directly determined
by the thrust ratio $\gamma\in\left(0,1\right)$ which is defined
as
\begin{equation}
\gamma\triangleq\frac{T_{\text{hover}}}{T_{\text{max}}}.\label{Eq31T0TpmaxThr}
\end{equation}
where thrust ratio $\gamma$ describes the remaining thrust for the
acceleration movement, which further determines the maximum forward
speed and the wind resistance ability of a multicopter. Therefore,
designers should estimate the desired $\gamma$ (usually $\gamma=0.5$
is selected for common multicopters) according to the kinematic performance
requirements of the multicopter. Then, the desired full-throttle thrust
$T_{\text{max}}$ of the propulsion system can be obtained through
Eqs.\,(\ref{eq:T0})(\ref{Eq31T0TpmaxThr}).

\subsection{Component Parameters}

The ultimate goal of the design optimization problem is to select
the optimal products from four component databases with component
parameters listed in Table\,\ref{tab1}, where $\Theta_{\text{p}}$,
$\Theta_{\text{m}}$, $\Theta_{\text{e}}$ and $\Theta_{\text{b}}$
represent the parameter sets for propellers, motors, ESCs and batteries.
In order to ensure commonality of the method, all component parameters
in Table\,\ref{tab1} are the basic parameters that can be easily
found in the product description pages. The detailed introduction
of each parameter in Table\,\ref{tab1} can also be found in \cite[pp. 31-46]{quan2017introduction}.

\begin{table}[ptbh]
\caption{Propulsion system parameters}
\label{tab1}\centering%
\begin{tabular}{|c|>{\raggedright}p{0.38\textwidth}|}
\hline 
Items  & Parameters \tabularnewline
\hline 
Propeller  & $\Theta_{\text{p}}\triangleq$\{Diameter $D_{\text{p}}$ (m), Pitch
Angle $\varphi_{\text{p}}$ (rad), Blade Number $B_{\text{p}}$\}\tabularnewline
\hline 
Motor  & $\Theta_{\text{m}}\triangleq$\{Nominal Maximum Voltage $U_{\text{mMax}}$
(V), Nominal Maximum Current $I_{\text{mMax}}$ (A), KV Value $K_{\text{V}}$
(RPM/V), No-load Current $I_{\text{m0}}$ (A), Resistance $R_{\text{m}}$
($\Omega$)\}\tabularnewline
\hline 
ESC  & $\Theta_{\text{e}}\triangleq$\{Nominal Maximum Voltage $U_{\text{eMax}}$
(V), Nominal Maximum Current $I_{\text{eMax}}$ (A), Resistance $R_{\text{e}}$
($\Omega$)\}; \tabularnewline
\hline 
Battery  & $\Theta_{\text{b}}\triangleq$\{Nominal Voltage $U_{\text{b}}$ (V),
Maximum Discharge Rate $K_{\text{b}}$ (A), Capacity $C_{\text{b}}$
(mAh), Resistance $R_{\text{b}}$ ($\Omega$)\}; \tabularnewline
\hline 
\end{tabular}
\end{table}

The propeller pitch angle $\varphi_{\text{p}}$ (unit: rad) in Table\,\ref{tab1}
is defined according to the propeller diameter ${D_{\text{{p}}}}$
(unit: m) and the propeller pitch ${H_{\text{{p}}}}$ (unit: m) as
\begin{equation}
\varphi_{\text{p}}\triangleq{\arctan\frac{{H_{\text{{p}}}}}{{\pi{D_{\text{{p}}}}}}}\label{Eq00-pitch}
\end{equation}
where ${H_{\text{{p}}}}$ and ${D_{\text{{p}}}}$ are usually contained
in the propeller model name. 

The most commonly used unit for the battery voltage $U_{\text{b}}$,
as well as the motor voltage $U_{\text{mMax}}$ and the ESC voltage
$U_{\text{eMax}}$, is ``S\textquotedblright , which denotes the
number of battery cells in series. For LiPo batteries, the voltage
changes from 4.2\,V to 3.7\,V as the battery capacity decreases
from full to empty, and the average voltage 4.0\,V is adopted for
the unit conversion. For example, $U_{\text{b}}=12\text{\,S}=48\text{\,V}.$

\subsection{Optimization Constraints}

\label{subsec:OptiConstr}

\subsubsection{Requirement Constraints}

According to \cite{Shi2017}, $t_{\text{hover}}$ and $T_{\text{max}}$
can be estimated by parameters $n_{\text{p}},T_{\text{hover}}$,$\Theta_{\text{p}},\Theta_{\text{m}},\Theta_{\text{e}},\Theta_{\text{b}}$.
Therefore, two equality constraints can be obtained according to the
design requirements in \textit{Section\,\ref{Sec2-Des}}.

\subsubsection{Safety Constraints}

The electric components should work within their allowed operating
conditions to prevent from being burnt out. Therefore, a series of
inequality constraints can be obtained for electric components including
the battery, the ESC and the motor of the propulsion system. The detailed
inequality expressions will be introduced in the later sections.

\subsubsection{Product Statistical Constraints}

In order to make sure that the obtained solution is meaningful and
practical, the product statistical features should be considered.
Otherwise, it is very possible that there is no product in reality
matching with the obtained component parameters. The product features
can be described by equality constraints based on the statistical
models of the products in the database. In practice, products with
different material, operating principle and processing technology
may have different product features, for example, LiPo batteries and
Ni-MH batteries. Therefore, different statistical models should be
obtained for different types of products, and designers should select
the required product type to the following optimization process to
improve the precision and practicability of the obtained result.

\subsection{Optimization Problem}

In practice, there are two methods to increase the flight time of
a multicopter: i) decrease the total weight of the multicopter to
allocate more free weight for the battery capacity; ii) increase the
efficiency of the multicopter propulsion system to decrease the required
battery capacity. For a typical multicopter, according to the weight
statistical model in \cite{Bershadsky2016a}, the propulsion system
weight is the main source (usually more than 70\%) of the multicopter
weight, and the weight of the battery is further the main source (usually
more than 60\%) of the propulsion system weight. In fact, for a propulsion
system, the maximum efficiency usually means the minimum battery weight
(capacity). Therefore, the above two methods essentially share the
same optimization objective, namely minimizing the weight of the propulsion
system.

Assuming that the total weight of the propulsion system is defined
as $G_{\text{sys}}$ (unit: N), the optimization objective of the
design optimization can be described as

\begin{equation}
\min_{\Theta_{\text{p}},\Theta_{\text{m}},\Theta_{\text{e}},\Theta_{\text{b}}}G_{\text{sys}}.\label{eq:OptiEqua}
\end{equation}
Along with the constraints in \textsl{Section\,\ref{subsec:OptiConstr}},
the optimal solutions for the parameters of each component can be
obtained from Eq.\,(\ref{eq:OptiEqua}).

\subsection{Problem Decomposition}

\label{subsec:ProbDeco}

According to \cite{Shi2017}, the detailed mathematical expressions
for Eq.\,(\ref{eq:OptiEqua}) are very complex because there are
15 parameters listed in Table\,\ref{tab1} that need to be optimized
through solving complex nonlinear equations. As a result, decomposition
and simplification are required to solve this problem.

\subsubsection{Weight Decomposition}

The total weight of the propulsion system $G_{\text{sys}}$ is determined
by the weight of each component as 
\begin{equation}
G_{\text{sys}}=n_{\text{p}}\left(G_{\text{p}}+G_{\text{m}}+G_{\text{e}}\right)+G_{\text{b}}\label{eq:Gsys}
\end{equation}
where $G_{\text{p}},G_{\text{m}},G_{\text{e}},G_{\text{b}}$ (unit:
N) denote the weight of the propeller, the motor, the ESC and the
battery respectively. Therefore, based on the idea of the greedy algorithm,
the optimization objective of minimizing $G_{\text{sys}}$ can be
decomposed into four sub-problems of minimizing the weight of each
component as 
\begin{equation}
\min G_{\text{sys}}\Rightarrow\min G_{\text{p}},\min G_{\text{m}},\min G_{\text{e}},\min G_{\text{b}}.\label{eq:WeightDecom}
\end{equation}
As mentioned above, the battery weight $G_{\text{b}}$ is the most
important factor for $G_{\text{sys}}$, and $G_{\text{b}}$ directly
depends on the battery capacity $C_{\text{b}}$. Since, according
to the analysis in \cite{quan2017introduction}, the battery capacity
$C_{\text{b}}$ depends on the efficiency of each component, the optimization
objective of minimizing $G_{\text{b}}$ can also be decomposed into
four sub-problems of maximizing the efficiency of each component as
\begin{equation}
\min G_{\text{b}}\Rightarrow\max\eta_{\text{p}},\max\eta_{\text{m}},\max\eta_{\text{e}},\max\eta_{\text{b}}.\label{eq:EffDecom}
\end{equation}
where $\eta_{\text{p}},\eta_{\text{m}},\eta_{\text{e}},\eta_{\text{b}}$
denote the efficiency of the propeller, the motor, the ESC and the
battery respectively.

\begin{figure}[tbh]
\centering \includegraphics[width=0.4\textwidth]{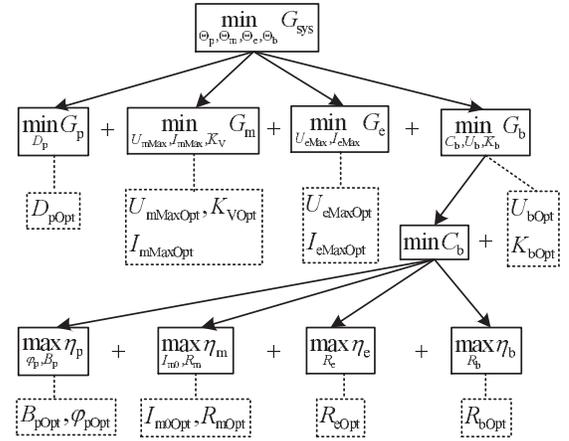} \caption{Optimization objective decomposition diagram.}
\label{Fig02Decom} 
\end{figure}

Thus, by solving the eight sub-problems in Eqs.\,(\ref{eq:WeightDecom})(\ref{eq:EffDecom}),
the optimal solutions for the parameters in Table\,\ref{tab1} can
be obtained with the results presented in Fig.\,\ref{Fig02Decom},
where the obtained optimal solutions are marked with a subscript ``Opt''.
Then, the optimal parameter sets for the propeller, the motor, the
ESC and the battery can be determined and represented by $\Theta_{\text{pOpt}}$,$\Theta_{\text{mOpt}}$,
$\Theta_{\text{eOpt}}$, and $\Theta_{\text{bOpt}}$ respectively.

\subsubsection{Product Selection Decomposition}

The ultimate goal of the design optimization of the propulsion system
is to determine the optimal propeller, motor, ESC and propeller products
from their corresponding databases according to the obtained optimal
parameter sets $\Theta_{\text{pOpt}},\Theta_{\text{mOpt}},\Theta_{\text{eOpt}},\Theta_{\text{bOpt}}$.
This problem can also be divided into four sub-problems. Through solving
the four sub-problems, the parameter sets of the obtained products
are represented by $\Theta_{\text{pOpt}}^{\ast}$, $\Theta_{\text{pOpt}}^{\ast}$,
$\Theta_{\text{pOpt}}^{\ast}$ and $\Theta_{\text{pOpt}}^{\ast}$.

\subsection{Solving Procedures}

\label{Sec2-End}

Through the above decomposition procedures, the whole design optimization
problem can finally be simplified, decoupled and divided into twelve
sub-problems. However, since there are argument-dependent relationships
among the sub-problems, the solving sequence should be well-arranged.
For instance, the propeller parameters $B_{\text{p}},\varphi_{\text{p}}$
are required for the motor optimization, and the optimal propeller
diameter $D_{\text{p}}$ depends on the obtained motor parameters.
As a result, the propeller and motor should be treated as a whole
during the solving procedures.

\begin{figure}[tbh]
\centering \includegraphics[width=0.4\textwidth]{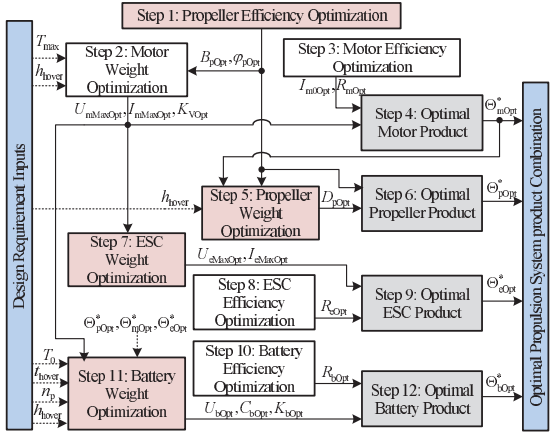} \caption{Solving procedures of the propulsion system optimization problem}
\label{Fig02-0} 
\end{figure}

In this paper, the twelve sub-problems will be solved separately by
twelve steps with the solving sequence shown in Fig.\,\ref{Fig02-0}.
Fig.\,\ref{Fig02-0} also presents the inputs, outputs, and parameter
dependency relationships of each sub-problem. The detailed solving
methods of each step will be introduced in \textit{Section\,\ref{Sec-4}}
with twelve subsections.

\section{Propulsion System Modeling}

\label{Sec-3}

\begin{figure}[tbh]
\centering \includegraphics[width=0.45\textwidth]{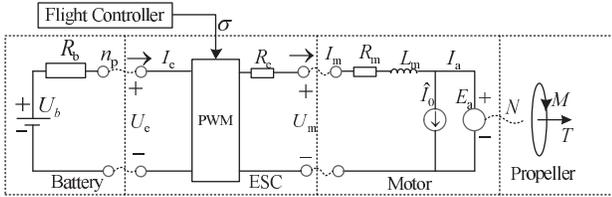} \caption{Equivalent circuit model of the whole propulsion system}
\label{Fig02} 
\end{figure}

The whole propulsion system can be modeled by the equivalent circuit
\cite{Shi2017} as shown in Fig.\,\ref{Fig02}, with which the optimization
sub-problems can be described by mathematical expressions.

\subsection{Propeller Modeling}

\subsubsection{Propeller Aerodynamic Model}

According to \cite{merchant2006propeller}, the thrust force $T$
(unit: N) and torque $M$ (unit: N$\cdot$m) of fixed-pitch propellers
can be obtained through equations 
\begin{equation}
\left\{ \begin{array}{c}
T=C_{\text{T}}\rho\left(\frac{N}{60}\right)^{2}D_{\text{p}}^{4}\\
M=C_{\text{M}}\rho\left(\frac{N}{60}\right)^{2}D_{\text{p}}^{5}
\end{array}\right.\label{Eq02PropTorque}
\end{equation}
where $N$ (unit: RPM) is the propeller revolving speed, $\rho$ is
the local air density (unit: kg/m$^{3}$), $C_{\text{T}}$ is the
propeller thrust coefficient, $C_{\text{M}}$ is the propeller torque
coefficient, and $D_{\text{p}}$ (unit: m, from ${\Theta_{\text{p}}}$)
is the propeller diameter. 

The air density $\rho$ is determined by the local temperature $T_{\text{t}}$
(unit: $^{\circ}$C) and the air pressure which is further determined
by the altitude $h_{\text{hover}}$ (unit: m). According to the international
standard atmosphere model \cite{cavcar2000international} 
\begin{equation}
\rho=f_{\rho}\left(h_{\text{hover}}\right)\triangleq\frac{273}{(273+T_{\text{t}})}(1-0.0065\frac{h_{\text{hover}}}{273+T_{\text{t}}})^{5.2561}\rho_{\text{0}}\label{Eq03AirDen}
\end{equation}
where the standard air density $\rho_{\text{0}}=1.293$kg/m$^{3}$
($^{\circ}$C, 273K).

The propeller coefficients $C_{\text{T}}$ and $C_{\text{M}}$ can
be modeled by using the blade element theory as presented in \cite{Shi2017}\cite{Mccrink2015}.
A simplified form is introduced here as 
\begin{equation}
\left\{ \begin{array}{lll}
C_{\text{T}} & =f_{C_{\text{T}}}\left(B_{\text{p}},D_{\text{p}},\varphi_{\text{p}}\right) & \triangleq k_{\text{t0}}B_{\text{p}}\varphi_{\text{p}}\\
C_{\text{M}} & =f_{C_{\text{M}}}\left(B_{\text{p}},D_{\text{p}},\varphi_{\text{p}}\right) & \triangleq k_{\text{m0}}B_{\text{p}}^{2}\left(k_{\text{m1}}+k_{\text{m2}}\varphi_{\text{p}}^{2}\right)
\end{array}\right.\label{Eq04Cm}
\end{equation}
where $k_{\text{t0}},k_{\text{m0}},k_{\text{m1}},k_{\text{m2}}$ are
constant parameters determined by the shapes and aerodynamic characteristics
of the propeller blades, and they can be obtained through the propeller
model in \cite{Shi2017} as 
\begin{equation}
\begin{array}{llcl}
k_{\text{t0}} & \triangleq\frac{0.25{\pi^{3}}\lambda{\zeta^{2}}{K_{0}}\varepsilon}{{\pi A+{K_{0}}}}, & k_{\text{m0}} & \triangleq\frac{1}{{8A}}{\pi^{2}}\lambda{\zeta^{2}}\\
k_{\text{m1}} & \triangleq C_{{\rm {fd}}}, & k_{\text{m2}} & \triangleq\frac{\pi AK_{0}^{2}\varepsilon}{e{{\left({\pi A+{K_{0}}}\right)}^{2}}}
\end{array}\label{eq:CTCMParamete}
\end{equation}
where the detailed definitions of the internal parameters of Eq.\,(\ref{eq:CTCMParamete})
can be found in \cite{Shi2017}. Note that $k_{\text{t0}},k_{\text{m0}},k_{\text{m1}},k_{\text{m2}}$
may slightly vary with the difference of types, material, and technology
of propellers. Based on the propeller data from T-MOTOR website \cite{TMotor2017},
general parameters $k_{\text{t0}},k_{\text{m0}},k_{\text{m1}},k_{\text{m2}}$
for the carbon fiber propellers are given by 
\begin{equation}
\begin{array}{ll}
k_{\text{t0}}=0.323, & k_{\text{m0}}=0.0432\\
k_{\text{m1}}=0.01, & k_{\text{m2}}=0.9
\end{array}.\label{Eq04KCM}
\end{equation}

\subsubsection{Propeller Efficiency Objective Function}

Similar to the lift-drag ratio for airfoils, a widely used efficiency
index to describe the efficiency of propellers is $\eta_{\text{T/M}}$,
which is defined as the ratio between the thrust coefficient $C_{\text{T}}$
and the torque coefficient $C_{\text{M}}$ as 
\begin{equation}
\eta_{\text{T/M}}\triangleq\frac{C_{\text{T}}}{C_{\text{M}}}=\frac{k_{\text{t0}}\varphi_{\text{p}}}{k_{\text{m0}}B_{\text{p}}\left(k_{\text{m1}}+k_{\text{m2}}\varphi_{\text{p}}^{2}\right)}\label{Eq05EffTM}
\end{equation}
where $\eta_{\text{T/M}}$ only depends on the aerodynamic design
of the blade shape, which is convenient for manufacturers to improve
the aerodynamic efficiency. Moreover, a higher $\eta_{\text{T/M}}$
means a smaller torque for generating the same thrust. Since $\eta_{\text{T/M}}$
is adopted by most of the manufacturers, this paper will use $\eta_{\text{T/M}}$
as the propeller efficiency objective function to obtain the optimal
$\varphi_{\text{pOpt}}$ and $B_{\text{pOpt}}$.

\subsubsection{Propeller Weight Objective Function}

Through analyzing the propeller products on the market, the propeller
weight $G_{\text{p}}$ can be described by a statistical model that
depends on the diameter $D_{\text{p}}$ and the blade number $B_{\text{p}}$
as 
\begin{equation}
G_{\text{p}}=f_{G_{\text{p}}}\left(B_{\text{p}},D_{\text{p}}\right)\label{eq:Gp}
\end{equation}
where $f_{G_{\text{p}}}\left(\cdot\right)$ is an increasing function
of $D_{\text{p}}$ and $B_{\text{p}}$. Therefore, the minimum propeller
weight $G_{\text{p}}$ requires that both $D_{\text{p}}$ and $B_{\text{p}}$
should be chosen as small as possible, which is described as 
\begin{equation}
\min G_{\text{p}}\Rightarrow\min B_{\text{p}},\min D_{\text{p}}.\label{eq:minGp}
\end{equation}

\subsection{Motor Modeling}

\subsubsection{Motor Circuit Model }

The equivalent circuit of a BLDC motor has been presented in Fig.\,\ref{Fig02},
where $U_{\text{m}}$ (unit: V) is the motor equivalent voltage and
$I_{\text{m}}$ (unit: A) is the motor equivalent current. According
to \cite{Shi2017,chapman2005electric}, $U_{\text{m}}$ and ${I_{\text{{m}}}}$
can be obtained through 
\begin{equation}
\left\{ \begin{array}{lll}
{I_{\text{{m}}}} & = & \frac{\pi{M{K_{\text{{V}}}}{U_{\text{{m0}}}}}}{{30({U_{\text{{m0}}}}-{I_{\text{{m0}}}}{R_{\text{{m}}}})}}+{I_{\text{{m0}}}}\\
{U_{\text{{m}}}} & = & {I_{\text{{m}}}R_{\text{{m}}}}+\frac{{{U_{\text{{m0}}}}-{I_{\text{{m0}}}}{R_{\text{{m}}}}}}{{{K_{\text{{V}}}}{U_{\text{{m0}}}}}}N
\end{array}\right.\label{Eq05Um}
\end{equation}
where $M$ (unit: N$\cdot$m) is the output torque of the motor which
equals to the propeller toque in Eq.\,(\ref{Eq02PropTorque}), $N$
is the motor rotating speed which equals to the propeller rotating
speed in Eq.\,(\ref{Eq02PropTorque}). The current and voltage measured
under no-load (no propeller) tests are called the no-load current
$I_{\text{{m0}}}$ (unit: A) and the no-load voltage $U_{\text{{m0}}}$
(unit: V), where ${U_{\text{{m0}}}}$ is a constant value defined
by manufacturers (usually ${U_{\text{{m0}}}=10}$\,V). Note that,
the nominal motor resistance $R_{\text{m0}}$ on the product description
is usually not accurate enough, so it is recommended to obtain the
actual resistance $R_{\text{m}}$ according to the test data of the
full-throttle current $I_{\text{{m}}}^{*}$ and speed $N^{*}$, where
a correction expression is derived from Eq.\,(\ref{Eq05Um}) as $R_{\text{m}}$$\approx\left(U_{\text{{b}}}-N^{*}/K_{\text{V}}\right)/I_{\text{{m}}}^{*}\approx2\sim3R_{\text{m0}}$
according to our experimental results. 

\subsubsection{Motor Constraints}

The actual operation voltage of the motor depends on the battery voltage
$U_{\text{b}}$ instead of the motor nominal maximum voltage (NMV)
$U_{\text{mMax}}$. The motor NMV $U_{\text{mMax}}$ defines the range
of the battery voltage $U_{\text{b}}$ that can ensure the motor work
safely, which is described as $U_{\text{b}}\leq U_{\text{mMax}}$.
According to \cite{Shi2017}, to prevent the motor from burnout, the
motor equivalent voltage $U$$_{\text{m}}$ and current $I_{\text{m}}$
satisfy the constraints that 
\begin{equation}
\begin{array}{l}
U_{\text{m}}\leq U_{\text{m\ensuremath{\sigma_{\text{max}}}}}=U_{\text{b}}\leq U_{\text{mMax}}\\
I_{\text{m}}\leq I_{\text{m\ensuremath{\sigma_{\text{max}}}}}\leq I_{\text{mMax}}
\end{array}\label{eq:MotConstraint}
\end{equation}
where $U_{\text{m\ensuremath{\sigma_{\text{max}}}}}$ and $I_{\text{m\ensuremath{\sigma_{\text{max}}}}}$
are the motor voltage and current under the full-throttle mode ($\sigma_{\text{max}}=1$).

By letting the motor work under the maximum limit condition as 
\begin{equation}
U_{\text{m}}={U_{\text{{mMax}}}}\text{, }I_{\text{m}}=I_{\text{mMax}}.\label{eq:MotorConstraint}
\end{equation}
Then, the maximum rotating speed ${N}_{\max}$ (unit: RPM) and torque
${M}_{\max}$ (unit: N$\cdot$m) of the motor can be obtained by combining
the motor model in Eq.\,(\ref{Eq05Um}) 
\begin{equation}
\left\{ \begin{array}{l}
{N}_{\max}=f_{{N}_{\max}}\left(\Theta_{\text{m}}\right)\triangleq\frac{{\left({U_{\text{{mMax}}}-{R_{\text{m}}}{I_{\text{mMax}}}}\right){K_{\text{V}}}{U_{\text{m0}}}}}{\left({{U_{\text{m0}}}-{I_{\text{m0}}}{R_{\text{m}}}}\right)}\\
{M}_{\max}=f_{{M}_{\max}}\left(\Theta_{\text{m}}\right)\triangleq\frac{{30\left({{I_{\text{mMax}}}-{I_{\text{m0}}}}\right)({U_{\text{m0}}}-{I_{\text{m0}}}{R_{\text{m}}})}}{\pi{K}_{\text{V}}{U_{\text{m0}}}}
\end{array}\right..\label{Eq18Mmax}
\end{equation}
Moreover, according to the propeller model in Eq.\,(\ref{Eq02PropTorque}),
the theoretical maximum thrust ${T_{\text{pMax}}}$ (unit: N) can
be obtained as 
\begin{equation}
T_{\text{pMax}}=\frac{{{C}_{\text{T}}}}{{{C}_{\text{M}}}}\frac{M_{\max}^{4/5}{{\rho}^{1/5}}{{C}_{\text{M}}^{1/5}{N}_{\max}^{2/5}}}{{{60}^{2/5}}}.\label{Eq19TpMax1}
\end{equation}
To satisfy the maneuverability requirement, the theoretical maximum
thrust range $\left[0,T_{\text{pMax}}\right]$ of the propulsion system
should cover the required thrust range $\left[0,T_{\text{max}}\right]$,
which means the following constraint should also be satisfied 
\begin{equation}
T_{\text{pMax}}\geq T_{\text{max}}.\label{eq:TmaxTpMaxEq}
\end{equation}

\begin{figure}[tbh]
\centering \includegraphics[width=0.4\textwidth]{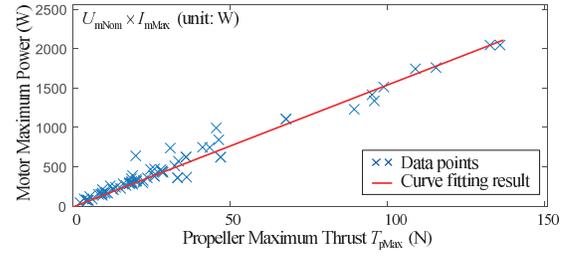} \caption{Statistical relationship between the motor maximum input power $U_{\text{{mMax}}}\cdot{I}_{\text{mMax}}$
and propeller maximum thrust $T_{\text{pMax}}$. The testing data
come from the motor products website \cite{TMotor2017}.}
\label{Fig05ThustWattVolt-1} 
\end{figure}

According to the statistical results in Fig.\,\ref{Fig05ThustWattVolt-1},
there is an equality constraint between the maximum input power ${U_{\text{{mMax}}}{I}_{\text{mMax}}}$
and the theoretical maximum thrust ${T_{\text{pMax}}}$ for motor
products as 
\begin{equation}
\frac{{T_{\text{pMax}}}}{{U_{\text{{mMax}}}{I}_{\text{mMax}}}}\approx G_{\text{WConst}}\label{Eq37GWconst}
\end{equation}
where $G_{\text{WConst}}$ (unit: N/W) is a constant coefficient that
reflects the technological process and product quality of products.
According to the curve fitting result in Fig.\,\ref{Fig05ThustWattVolt-1},
the coefficient for the tested motors is $G_{\text{WConst}}\approx0.0624$.

\subsubsection{Motor Efficiency Objective Function }

The motor power efficiency $\eta_{\text{m}}$ is defined as 
\begin{equation}
\eta_{\text{m}}\triangleq\frac{P_{\text{p}}}{P_{\text{m}}}=\frac{M\frac{2\pi N}{60}}{U_{\text{m}}I_{\text{m}}}.
\end{equation}
According to Eq.\,(\ref{Eq05Um}), the propeller torque $M$ and
rotating speed $N$ can be described by $U_{\text{m}}\text{ and }I_{\text{m}}$,
which yields that 
\begin{equation}
\begin{array}{c}
\eta_{\text{m}}=\left(1-\frac{I_{\text{m}}}{{U_{\text{{m}}}}}{R_{\text{{m}}}}\right)\left(1-\frac{1}{I_{\text{m}}}I_{\text{m0}}\right)\end{array}.\label{eq:EqEffmOtor}
\end{equation}

It can be observed from Eq.\,(\ref{eq:EqEffmOtor}) that the motor
efficiency $\eta_{\text{m}}$ has negative correlations with $R_{\text{{m}}}$
and $I_{\text{m0}}$. Therefore, $R_{\text{{m}}}$ and $I_{\text{m0}}$
should be chosen as small as possible for the maximum motor efficiency,
which is described as 
\begin{equation}
\max\eta_{\text{m}}\Rightarrow\min R_{\text{{m}}},\min I_{\text{m0}}.\label{eq:EffM}
\end{equation}

\subsubsection{Weight Optimization Objective Function}

Through analyzing the motor products on the market, the motor weight
$G_{\text{m}}$ can be described by a statistical model depending
on the motor nominal maximum voltage (NMV) $U_{\text{{mMax}}}$, the
Nominal Maximum Current (NMC) $I_{\text{mMax}}$, and the KV value
$K_{\text{V}}$ as 
\begin{equation}
G_{\text{m}}=f_{G_{\text{m}}}\left(U_{\text{mMax}},I_{\text{mMax}},K_{\text{V}}\right).\label{eq:GmOrgin}
\end{equation}
According to Eqs.\,(\ref{Eq19TpMax1})(\ref{Eq37GWconst}), $I_{\text{mMax}}$
and $K_{\text{V}}$ can be described by $U_{\text{mMax}}$ and $T_{\text{pMax}}$.
Therefore, Eq.\,(\ref{eq:GmOrgin}) can be rewritten into the following
form 
\begin{equation}
G_{\text{m}}=f_{G_{\text{m}}}^{\prime}\left(U_{\text{mMax}},T_{\text{pMax}}\right).\label{eq:Gm}
\end{equation}
Thus, by combining the constraints in Eqs.\,(\ref{eq:MotConstraint})(\ref{eq:TmaxTpMaxEq}),
the motor weight optimization problem can be written into 
\begin{equation}
\begin{array}{c}
\underset{U_{\text{mMax}},T_{\text{pMax}}}{\min}f_{G_{\text{m}}}^{\prime}\left(U_{\text{mMax}},T_{\text{pMax}}\right)\\
\text{s.t. }U_{\text{b}}\leq U_{\text{mMax}},T_{\text{max}}\leq T_{\text{pMax}}
\end{array}.\label{eq:minGm}
\end{equation}

In practice, the motor weight $G_{\text{m}}$ has a positive correlation
with $U_{\text{mMax}}\text{ and }T_{\text{pMax}}$. Therefore, the
minimum motor weight $G_{\text{m}}$ requires that the $U_{\text{mMax}}$
and $T_{\text{pMax}}$ should both be chosen as small as possible,
which is described as 
\begin{equation}
\min G_{\text{m}}\Rightarrow\min U_{\text{mMax}},\min T_{\text{pMax}}.\label{eq:minGp-1}
\end{equation}
Thus, solving Eq.\,(\ref{eq:minGm}) gives that 
\begin{equation}
U_{\text{b}}=U_{\text{mMax}},T_{\text{max}}=T_{\text{pMax}}\label{eq:UbTmax}
\end{equation}
where $U_{\text{mMax}}$ should be chosen as small as possible.

\subsection{ESC Modeling}

\subsubsection{ESC Circuit Model}

After receiving the throttle signal $\sigma{\in}\left[0,1\right]$
from the flight controller, the ESC converts the direct-current power
of the battery to the PWM-modulated voltage $\sigma U_{\text{{e}}}$
for the BLDC motor without speed feedback. Then, the motor rotating
speed is determined by both the motor and propeller models. Since
the propeller model is nonlinear, the motor speed is not in proportion
to the input throttle signal. According to the ESC equivalent circuit
in Fig.\,\ref{Fig02}, the ESC current $I_{\text{e}}$ (unit: A)
and voltage $U_{\text{e}}$ (unit: V) are given by 
\begin{equation}
\begin{array}{cl}
\sigma U_{\text{{e}}} & =U_{\text{{m}}}+I_{\text{{m}}}R_{\text{e}}\\
I_{\text{{e}}} & ={\sigma I_{\text{{m}}}}
\end{array}.\label{Eq06Ie}
\end{equation}

\subsubsection{ESC Efficiency Objective Function}

According to Eq.\,(\ref{Eq06Ie}), the power efficiency of the ESC
can be obtained as 
\begin{equation}
\begin{array}{lll}
\eta_{\text{e}} & \triangleq & \frac{U_{\text{m}}I_{\text{m}}}{U_{\text{e}}I_{\text{e}}}=\frac{1}{1+\frac{I_{\text{m}}}{U_{\text{m}}}R_{\text{e}}}\end{array}
\end{equation}
which shows that the ESC efficiency $\eta_{\text{e}}$ increases as
the resistance $R_{\text{e}}$ decreases, which yields that 
\begin{equation}
\max\eta_{\text{e}}\Rightarrow\min R_{\text{e}}.\label{eq:EffE}
\end{equation}

\subsubsection{ESC Weight Objective Function}

Through analyzing the ESC products on the market, the ESC weight $G_{\text{e}}$
can be described by a statistical model depending on the ESC nominal
maximum voltage (NMV) $U_{\text{{eMax}}}$, the Nominal Maximum Current
(NMC) $I_{\text{eMax}}$ as 
\begin{equation}
G_{\text{e}}=f_{G_{\text{e}}}\left(U_{\text{eMax}},I_{\text{eMax}}\right)\label{eq:GeOrgin}
\end{equation}
where $f_{G_{\text{e}}}\left(\cdot\right)$ is in positive correlation
with $U_{\text{eMax}}\text{ and }I_{\text{eMax}}$. Therefore, to
minimize $G_{\text{e}}$, the ESC parameters $U_{\text{eMax}},I_{\text{eMax}}$
should be chosen as small as possible, which is described as

\begin{equation}
\min G_{\text{e}}\Rightarrow\min U_{\text{eMax}},\min I_{\text{eMax}}.\label{eq:GeMin}
\end{equation}

\subsection{Battery Modeling}

\subsubsection{Battery Circuit Model}

The battery is used to provide energy to drive the motor through ESC.
The most commonly-used type battery is the LiPo battery because of
the superior performance and low price. According to Fig.\,\ref{Fig02},
the battery model is given by 
\begin{equation}
U_{\text{b}}=U_{\text{\ensuremath{\text{e}}}}+{I}_{\text{b}}{R}_{\text{b}}\label{Eq08Ue}
\end{equation}
where ${U}_{\text{b}}$ (unit: V) is the nominal battery voltage,
and ${I}_{\text{b}}$ (unit: A) is the output current.

Assuming that the number of the propulsion unit on a multicopter is
$n_{\text{p}}$, the battery current is given by 
\begin{equation}
{I_{\text{{b}}}=n_{\text{p}}I_{\text{{e}}}+I_{\text{other}}}\label{Eq09Ob}
\end{equation}
where $I_{\text{other}}$ (unit: A) is the current from other devices
on the multicopter such as the flight controller and the camera. Usually,
according to \cite{quan2017introduction}, it can be assumed that
$I_{\text{other}}\approx0.5\text{\,A}$ if there is only a flight
controller on the multicopter.

According to \cite{Shi2017}, the battery discharge time $t_{\text{discharge}}$
(unit: min) is determined by the battery capacity $C_{\text{b}}$
and the discharge current $I_{\text{b}}$ 
\begin{equation}
t_{\text{discharge}}\approx\frac{0.85C_{\text{b}}}{I_{\text{b}}}\cdot\frac{60}{1000}\label{eq:tdischarge}
\end{equation}
where the coefficient 0.85 denotes a 15\% remaining capacity to avoid
over discharge. Note that the endurance computation equations in Eqs.\,(\ref{Eq08Ue})-(\ref{eq:tdischarge})
are simplified to reduce the computation time. They can be replaced
by more accurate and nonlinear methods as presented in \cite{donateo2017design,donateo2017new}
to increase the precision of the battery optimization result.

\subsubsection{Battery Constraints}

The Maximum Discharge Rate (MDR) $K_{\text{b}}$ (unit: mA/mAh or
marked with symbol ``C'') of the battery is defined as 
\begin{equation}
K_{\text{b}}=\frac{1000I_{\text{bMax}}}{C_{\text{b}}}
\end{equation}
where $I_{\text{bMax}}$ (unit: A) is the maximum discharge current
that the battery can withstand. Since the battery should be able to
work safely under the full-throttle mode of the motor, the maximum
discharge current $I_{\text{bMax}}$ should satisfy 
\begin{equation}
I_{\text{bMax}}\geq n_{\text{p}}I_{\text{e\ensuremath{\sigma_{\text{max}}}}}+I_{\text{other}}=n_{\text{p}}I_{\text{{mMax}}}+I_{\text{other}}\label{eq:Kbbb}
\end{equation}
which yields that 
\begin{equation}
K_{\text{b}}\geq\frac{1000\left(n_{\text{p}}I_{\text{{mMax}}}+I_{\text{other}}\right)}{C_{\text{b}}}.\label{eq:KbConstraint}
\end{equation}

\subsubsection{Battery Efficiency Objective Function}

According to Eqs.\,(\ref{Eq08Ue})(\ref{Eq09Ob}), the battery efficiency
$\eta_{\text{b}}$ can be written into 
\begin{equation}
\begin{array}{cc}
\eta_{\text{b}} & \triangleq\frac{n_{\text{p}}U_{\text{e}}I_{\text{e}}}{U_{\text{b}}I_{\text{b}}}=\left(1-\frac{I_{\text{b}}}{U_{\text{b}}}R_{\text{b}}\right)\left(1-\frac{I_{\text{other}}}{I_{\text{b}}}\right)\end{array}\label{Eq09BEFF}
\end{equation}
which shows that the battery efficiency $\eta_{\text{b}}$ increases
as the resistance $R_{\text{b}}$ decreases, which is described as

\begin{equation}
\max\eta_{\text{b}}\Rightarrow\min R_{\text{b}}.\label{eq:Effb}
\end{equation}

\subsubsection{Battery Weight Objective Function}

According to the definition of the battery power density $\rho_{\text{b}}$
(unit: Wh/kg), the battery weight can be described as 
\begin{equation}
G_{\text{b}}=\frac{C_{\text{b}}U_{\text{b}}}{1000g\rho_{\text{b}}}\label{eq:Gb}
\end{equation}
where $g=9.8\text{\,m/s\ensuremath{^{2}}}$ is the acceleration of
gravity. Limited by the battery technology, the power density $\rho_{\text{b}}$
for a specific battery type is statistically close to a constant value.
For instance, $\rho_{\text{b}}\text{\ensuremath{\approx}}140\text{\,Wh/kg}$
for LiPo batteries. Moreover, according to the statistical results,
the battery weight is positively correlated with $K_{\text{b}}$.
Therefore, to minimize $G_{\text{b}}$, the battery parameters $C_{\text{b}}$,
$U_{\text{b}}$ and $K_{\text{b}}$should be chosen as small as possible,
which is described as

\begin{equation}
\min G_{\text{b}}\Rightarrow\min U_{\text{b}},\min K_{\text{b}},\min C_{\text{b}}.
\end{equation}

\section{Design Optimization}

\label{Sec-4}

\subsection{Step 1: Propeller Efficiency Optimization}

\label{Sec-DesOpProp}

\subsubsection{Optimal Blade Number $B_{\text{pOpt}}$}

According to Eq.\,(\ref{Eq05EffTM}), if only the blade number $B_{\text{p}}$
is considered, the propeller thrust efficiency $\eta_{\text{T/M}}$
can be simplified into the following form 
\begin{equation}
\eta_{\text{T/M}}\propto\frac{1}{B_{\text{p}}}\label{Eqw14-2Bp}
\end{equation}
where the symbol ``$\propto$'' means ``in proportion to''. Eq.\,(\ref{Eqw14-2Bp})
indicates that $\eta_{\text{T/M}}$ monotonically decreases as ${B_{\text{{p}}}}$
increases. Therefore, to maximize $\eta_{\text{T/M}}$, the blade
number ${B_{\text{{p}}}}$ should be chosen as small as possible.
Moreover, according to the weight optimization principle in Eq.\,(\ref{eq:minGp}),
the blade number should also be chosen as small as possible to minimize
the propeller weight $G_{\text{p}}$. Considering that the blade number
should satisfy the constraint that ${B_{\text{{p}}}\geq2}$, the optimal
blade number ${B_{\text{{pOpt}}}}$ should be chosen as 
\begin{equation}
{B_{\text{{pOpt}}}=2}.\label{Eq14BpOpt}
\end{equation}

\subsubsection{Optimal Pitch Angle $\varphi_{\text{pOpt}}$}

According to Eq.\,(\ref{Eq05EffTM}), if only the pitch angle $\varphi_{\text{p}}$
is considered, $\eta_{\text{T/M}}$ can be simplified into the form
as 
\begin{equation}
\eta_{\text{T/M}}\propto\frac{1}{k_{\text{m1}}\varphi_{\text{p}}^{-1}+k_{\text{m2}}\varphi_{\text{p}}}.\label{EqfH14EffFHD}
\end{equation}
It is easy to obtain from Eq.\,(\ref{EqfH14EffFHD}) that $\eta_{\text{T/M}}$
first increases then decreases as $\varphi_{\text{p}}$ increases.
Therefore, the optimal pitch angle $\varphi_{\text{pOpt}}$ should
be able to maximize $\eta_{\text{T/M}}$, which yields that 
\begin{equation}
\varphi_{\text{pOpt}}=\sqrt{\frac{k_{\text{m1}}}{k_{\text{m2}}}}.\label{Eq15DHOpt0}
\end{equation}

Noteworthy, if it happens in some cases that there are few propeller
products in database with pitch angles close to the obtained $\varphi_{\text{pOpt}}$,
then the mean pitch angle $\overline{\varphi}_{\text{p}}$ can be
chosen as the optimal pitch angle $\varphi_{\text{pOpt}}$=$\overline{\varphi}_{\text{p}}$
to ensure the method can find a proper propeller product.

\subsection{Step 2: Motor Weight Optimization}

\label{subsec:motorWeight}

\subsubsection{Optimal Motor NMV $U_{\text{mMaxOpt}}$}

\begin{figure}[tbh]
\centering \includegraphics[width=0.4\textwidth]{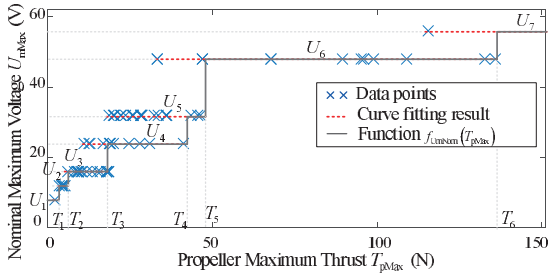} \caption{Statistical relationship between the motor NMV $U_{\text{{mMax}}}$
and the propeller maximum thrust $T_{\text{pMax}}$. The testing data
come from the motor products website \cite{TMotor2017}.}
\label{Fig05ThustWattVolt} 
\end{figure}

According to statistical results in Fig.\,\ref{Fig05ThustWattVolt},
the relationship between ${U_{\text{{mMax}}}}$ and ${T_{\text{pMax}}}$
can be described as a piecewise function 
\begin{equation}
f{_{{U_{\text{{mMax}}}}}}\left({T_{\text{pMax}}}\right){\triangleq}\left\{ \begin{array}{cc}
U_{1}, & {0}<{T_{\text{pMax}}}\leq{T}_{1}\\
U_{2}, & {T}_{1}<{T_{\text{pMax}}}\leq{T}_{2}\\
\cdots & \cdots
\end{array}\right.\label{Eq36UmMaxFunc}
\end{equation}
where $f{_{{U_{\text{{mMax}}}}}}\left(\cdot\right)$ is obtained with
the principle of minimizing $U_{\text{{mMax}}}$ for a given $T_{\text{pMax}}$.

Thus, by substituting $T_{\text{pMax}}=T_{\text{max}}$ into Eq.\,(\ref{Eq36UmMaxFunc}),
the optimal motor NMV $U_{\text{{mMaxOpt}}}$ can be obtained as 
\begin{equation}
{U_{\text{{mMaxOpt}}}=f{_{{U_{\text{{mMax}}}}}}\left({T_{\text{max}}}\right)}.\label{Eq36UmMax}
\end{equation}

\subsubsection{Optimal NMC $I_{\text{{mMaxOpt}}}$}

According to Eq.\,(\ref{Eq37GWconst}), there is a function relationship
between $T_{\text{pMax}}$, $I_{\text{{mMax}}}$ and $U_{\text{{mMax}}}$.
Therefore, by substituting $U_{\text{{mMax}}}=U_{\text{{mMaxOpt}}}$
and $T_{\text{pMax}}=T_{\text{max}}$ into Eq.\,(\ref{Eq37GWconst}),
the optimal motor NMC $I_{\text{{mMaxOpt}}}$ can be obtained as 
\begin{equation}
{I_{\text{{mMaxOpt}}}=\frac{{T_{\text{max}}}}{G_{\text{WConst}}U_{\text{{mMaxOpt}}}}}.\label{Eq37ImMaX1}
\end{equation}

\subsubsection{Optimal KV Value ${K}_{\text{VOpt}}$}

Since the motor no-load current ${I_{\text{m0}}}$ and resistance
$R_{\text{m}}$ are both very small in practice, it is reasonable
to assume that 
\begin{equation}
{I_{\text{m0}}}\approx0\text{ and }{R_{\text{m}}\approx0}.\label{Eq33ImRm}
\end{equation}
Substituting Eq.\,(\ref{Eq33ImRm}) into Eq.\,(\ref{Eq18Mmax})
gives 
\begin{equation}
\begin{array}{l}
M_{\max}\approx\frac{30{{I}_{\text{mMax}}}}{\pi{{K}_{\text{V}}}}\\
N_{\max}\approx{U_{\text{{mMax}}}{K}_{\text{V}}}
\end{array}.\label{Eq34MmNm}
\end{equation}
Thus, Eq.\,(\ref{Eq19TpMax1}) can be simplified into the following
form 
\begin{equation}
{T_{\text{pMax}}\approx{k}_{\text{tm}}}\cdot{{\left(\frac{{{I}_{\text{mMax}}^{2}U_{\text{{mMax}}}}}{{{K}_{\text{V}}}}\right)}^{2/5}}\label{EQ35IUK}
\end{equation}
where ${{k}_{\text{tm}}}$ is defined as 
\begin{equation}
k_{\text{tm}}=f_{k_{\text{tm}}}\left(\text{\ensuremath{B_{\text{p}},\varphi_{\text{p}},h_{\text{hover}}}}\right)\triangleq\sqrt[5]{k_{\text{c}}\frac{255\rho C_{\text{T}}^{5}}{\pi^{4}C_{\text{M}}^{4}}}\label{Eq36Km}
\end{equation}
where $\rho=f_{\rho}\left(h_{\text{hover}}\right),$ $C_{\text{T}}=f_{C_{\text{T}}}\left(B_{\text{p}},D_{\text{p}},\varphi_{\text{p}}\right)$
and $C_{\text{M}}=f_{C_{\text{M}}}\left(B_{\text{p}},D_{\text{p}},\varphi_{\text{p}}\right)$
are defined in Eqs.\,(\ref{Eq03AirDen})(\ref{Eq04Cm}), and $k_{\text{c}}$
is a constant correction coefficient to compensate for the neglected
factors including $I_{\text{m0}}$, $R_{\text{m}}$. According to
the statistical analysis, the correction coefficient can be set as
$k_{\text{c}}$$\approx0.82$$.$

Finally, by substituting $U_{\text{{mMaxOpt}}},$ $I_{\text{mMaxOpt}}$,
$T_{\text{max}}$, $B_{\text{pOpt}},$ $\varphi_{\text{pOpt}}$ and
$h_{\text{hover}}$ into Eqs.\,(\ref{EQ35IUK})(\ref{Eq36Km}), the
expression for the optimal KV value ${{K}_{\text{VOpt}}}$ is obtained
as 
\begin{equation}
{K}_{\text{VOpt}}=f_{{k}_{\text{tm}}}^{5/2}\left(\text{\ensuremath{B_{\text{pOpt}},\varphi_{\text{pOpt}},h_{\text{hover}}}}\right)\frac{{{I}_{\text{mMaxOpt}}^{2}U_{\text{{mMaxOpt}}}}}{{T_{\text{max}}^{5/2}}}.\label{Eq37KVOpt}
\end{equation}

\subsection{Step 3: Motor Efficiency Optimization}

\subsubsection{Optimal Motor Resistance ${R_{\text{{m}Opt}}}$ and No-load Current
$I_{\text{m0Opt}}$}

In practice, $R_{\text{{m}}}$ and $I_{\text{m0}}$ should satisfy
the constraint that 
\begin{equation}
R_{\text{{m}}}>0\text{ and }I_{\text{m0}}>0.\label{eq:mRE}
\end{equation}
By combining Eq.\,(\ref{eq:EffM}) and Eq.\,(\ref{eq:mRE}), the
optimal motor resistance ${R_{\text{{m}Opt}}}$ and no-load current
$I_{\text{m0Opt}}$ are marked with 
\begin{equation}
{R_{\text{{m}Opt}}=0}\text{ and }I_{\text{m0Opt}}=0
\end{equation}
which denote that the motor resistance and no-load current should
be chosen as close to zero as possible. 

Note that, in addition to parameters $R_{\text{{m}}}$,$I_{\text{m0}}$,
the motor efficiency $\eta_{\text{m}}$ in Eq. (\ref{eq:EqEffmOtor})
is also determined by the motor working state ($U_{\text{m}}$,$I_{\text{m}}$)
which further depends on factors including the throttle, rotating
speed, propeller parameters, other motor parameters, air density,
etc. According to our experimental results, a larger motor with a
larger propeller will have a higher efficiency for generating the
same thrust. Therefore, only maximizing the motor efficiency $\eta_{\text{m}}$
to solve all the motor parameters may not obtain the desired result
because the size and weight will be very large. Thus, this paper finds
the optimal propulsion system by considering both the efficiency and
the weight to ensure the practicability of the obtained results.

\subsection{Step 4: Optimal Motor Product}

\label{SecOptMotSel}

Although the optimal motor parameters $\Theta_{\text{mOpt}}\triangleq\{U_{\text{mMaxOpt}}$,
$I_{\text{mMaxOpt}}$, $K_{\text{VOpt}}$, $R_{\text{mOpt}}$, $I_{\text{m0Opt}}\}$
have been obtained through the above procedures, it is still difficult
to determine a corresponding product from the database. For example,
$U_{\text{mMax}}$ of motor products are usually given by discrete
form like 20A, 30A, 40A, while the obtained solutions are usually
given with continuous form like $I_{\text{mMaxOpt}}=33.5$\,A. To
solve this problem, a method is proposed to determine the optimal
motor product according to $\Theta_{\text{mOpt}}$. For simplicity,
the parameter set of the obtained motor product is represented by
$\Theta_{\text{mOpt}}^{\ast}\triangleq\{U_{\text{mMaxOpt}}^{\ast}$,
$I_{\text{mMaxOpt}}^{\ast}$, $K_{\text{VOpt}}^{\ast}$, $R_{\text{mOpt}}^{\ast}$,
$I_{\text{m0Opt}}^{\ast}\}$.

There are two selection principles for the optimal motor product:

(i) The product should be selected by comparing with every parameter
of $\Theta_{\text{mOpt}}$ in a proper sequence. Through the statistical
analysis of the motor products on the market, a comparison sequence
is given by considering the influence of each parameter on the motor
weight as 
\begin{equation}
U_{\text{mMaxOpt}}^{*}\text{, }K_{\text{VOpt}}^{*}\text{, }I_{\text{mMaxOpt}}^{*}\text{, }R_{\text{{mOpt}}}^{*}\text{, }{I_{\text{m0Opt}}^{*}}.\label{eq:secPrin}
\end{equation}

(ii) When comparing one parameter, on the premise of ensuring safety
requirements, the product should be chosen equal to or close to the
corresponding parameter in $\Theta_{\text{mOpt}}$. For example, if
$I_{\text{mMaxOpt}}=33.5$\,A and the available current options are
20A, 30A, 40A, then it should be chosen that $I_{\text{mMaxOpt}}^{\ast}=40$\,A
for some safety margin. The safety constraints for the selection of
motor products are given by 
\begin{equation}
\begin{array}{l}
U_{\text{mMaxOpt}}^{\ast}\geq U_{\text{mMaxOpt}},\,I_{\text{mMaxOpt}}^{\ast}\geq I_{\text{mMaxOpt}}\end{array}.\label{EqMotorSeq}
\end{equation}

\subsection{Step 5: Propeller Weight Optimization}

\subsubsection{Optimal Diameter $D_{\text{pOpt}}$}

According to Eq.\,(\ref{eq:TmaxTpMaxEq}), the following constraint
equation should be satisfied 
\begin{equation}
T_{\text{pMax}}=C_{\text{T}}\rho\left(\frac{N_{\text{max}}}{60}\right)^{2}D_{\text{p}}^{4}\geq T_{\text{max}}
\end{equation}
which yields that 
\begin{equation}
D_{\text{p}}\geq\sqrt[4]{\frac{60^{2}T_{\text{max}}}{C_{\text{T}}\rho N_{\text{max}}^{2}}}.\label{eq:MinDia}
\end{equation}

According to the optimization objective in Eq.\,(\ref{eq:minGp}),
the optimal diameter should be chosen as the minimum diameter under
constraint in Eq.\,(\ref{eq:MinDia}). Therefore, the optimal diameter
$D_{\text{pOpt}}$ can be obtained by combining Eqs.\,(\ref{Eq03AirDen})(\ref{Eq04Cm})(\ref{Eq18Mmax})
with the parameters $h_{\text{hover}}$, $D_{\text{pOpt}}$, $B_{\text{pOpt}}$
and $\Theta_{\text{mOpt}}^{*}$ as 
\begin{equation}
\begin{array}{cl}
D_{\text{pOpt}} & =\sqrt[4]{\frac{60^{2}T_{\text{max}}}{\rho C_{\text{T}}N_{\text{max}}^{2}}}=\sqrt[4]{\frac{60^{2}T_{\text{pMax}}}{\rho C_{\text{T}}N_{\text{max}}^{2}}}=\sqrt[5]{{\frac{{M}_{\max}}{{\rho{C_{\text{M}}}{{\left({\frac{{N}_{\max}}{{60}}}\right)}^{2}}}}}}\\
 & =\sqrt[5]{{\frac{3600f_{{M}_{\max}}\left(\Theta_{\text{mOpt}}^{*}\right)}{f_{\rho}\left(h_{\text{hover}}\right)f_{C_{\text{M}}}\left(B_{\text{pOpt}},D_{\text{pOpt}}\right)f_{N_{\text{max}}}^{2}\left(\Theta_{\text{mOpt}}^{*}\right)}}}.
\end{array}\label{Eq19DpOpt}
\end{equation}

Since the propeller pitch $H_{\text{{p}}}$ is more convenient to
select a propeller product, according to the definition of the pitch
angle $\varphi_{\text{p}}$ in Eq.\,(\ref{Eq00-pitch}), the optimal
propeller pitch $H_{\text{{pOpt}}}$ is given by
\[
H_{\text{{pOpt}}}=\pi\cdot D_{\text{pOpt}}\cdot\tan\varphi_{\text{pOpt}}.
\]

\subsection{Step 6: Optimal Propeller Product}

\label{SecOpProp}

With the obtained parameters $\Theta_{\text{pOpt}}=\left\{ B_{\text{pOpt}},\varphi_{\text{p}},D_{\text{pOpt}}\right\} $,
the optimal propeller product can be determined by searching the propeller
product database. The parameter set of the obtained optimal propeller
product is represented by $\Theta_{\text{pOpt}}^{\ast}=\left\{ B_{\text{pOpt}}^{*},\varphi_{\text{p}}^{*},D_{\text{pOpt}}^{*}\right\} $.

Similar to the selection principles in \textit{Section\,\ref{SecOptMotSel}},\textit{
}the optimal propeller product should be determined from the propeller
database by comparing the parameters in the sequence $B_{\text{pOpt}}^{*}$,$\varphi_{\text{p}}^{*}$,$D_{\text{pOpt}}^{*}$,
and each parameter in $\Theta_{\text{pOpt}}^{\ast}$ should be chosen
equal or close to the corresponding parameter in $\Theta_{\text{pOpt}}$
with satisfying the safety constraints 
\begin{equation}
\begin{array}{c}
B_{\text{pOpt}}^{*}=B_{\text{pOpt}},\,D_{\text{pOpt}}^{*}\leq D_{\text{pOpt}}\end{array}.\label{eq:DpOpt}
\end{equation}

\subsection{Step 7: ESC Weight Optimization}

\subsubsection{Optimal ESC NMV $U_{\text{eMaxOpt}}$ and NMI $I_{\text{eMaxOpt}}$}

Since the ESC and the motor are connected in series, their voltage
and current should match with each other to ensure proper operations.
For the safety, the ESC NMV $U_{\text{eMax}}$ and NMI $I_{\text{eMax}}$
should be able to support the maximum voltage and current from the
motor, which means 
\begin{equation}
\begin{array}{c}
U_{\text{eMax}}\geq U_{\text{mMaxOpt}},\,I_{\text{eMax}}\geq I_{\text{mMaxOpt}}\end{array}.\label{eq:GeCons}
\end{equation}
By combining Eq.\,(\ref{eq:GeMin}) and Eq.\,(\ref{eq:GeCons}),
the optimal $U_{\text{eMaxOpt}}$ and $I_{\text{eMaxOpt}}$ are given
by 
\begin{equation}
\begin{array}{c}
U_{\text{eMaxOpt}}=U_{\text{mMaxOpt}},\,I_{\text{eMaxOpt}}=I_{\text{mMaxOpt}}\end{array}.\label{eq:OptESC}
\end{equation}

\subsection{Step 8: ESC Efficiency Optimization}

\subsubsection{Optimal ESC Resistance $R_{\text{eOpt}}$}

According to Eq.\,(\ref{eq:EffE}), the optimal ESC resistance $R_{\text{eOpt}}$
should be chosen as small as possible for the maximum ESC efficiency
$\eta_{\text{e}}$. Since $R_{\text{e}}>0$ in practice, the optimal
ESC resistance $R_{\text{eOpt}}$ is marked with 
\begin{equation}
R_{\text{eOpt}}=0
\end{equation}
which denotes that the ESC resistance should be chosen as close to
zero as possible.

\subsection{Step 9: Optimal ESC Product}

With the obtained parameters $U_{\text{eMaxOpt}},I_{\text{eMaxOpt}},R_{\text{eOpt}}$,
the optimal ESC product can be determined by searching the ESC product
database. The parameter set of the obtained optimal ESC product is
represented by $\Theta_{\text{eOpt}}^{\ast}=\left\{ U_{\text{eMaxOpt}}^{*},I_{\text{eMaxOpt}}^{*},R_{\text{eOpt}}^{*}\right\} $.

Similar to the selection principles in \textit{Section\,\ref{SecOptMotSel}},
the optimal ESC product should be determined from the ESC database
by comparing the parameters in the sequence $U_{\text{eMaxOpt}}^{*},I_{\text{eMaxOpt}}^{*},R_{\text{eOpt}}^{*}$,
and each parameter in $\Theta_{\text{eOpt}}^{\ast}$ should be chosen
equal or close to the corresponding parameter in $\Theta_{\text{eOpt}}$
with satisfying the constraints 
\begin{equation}
\begin{array}{c}
U_{\text{eMaxOpt}}^{\ast}\geq U_{\text{eMaxOpt}},\,I_{\text{eMaxOpt}}^{\ast}\geq I_{\text{eMaxOpt}}\end{array}.
\end{equation}

\subsection{Step 10: Battery Efficiency Optimization}

\subsubsection{Optimal Battery Resistance ${R}_{\text{bOpt}}$}

According to Eq.\,(\ref{eq:Effb}), $R_{\text{b}}$ should be chosen
as small as possible for the maximum battery efficiency $\eta_{\text{b}}$.
considering that $R_{\text{b}}>0$ in practice, the optimal battery
resistance ${R}_{\text{bOpt}}$ is marked with 
\begin{equation}
{R}_{\text{bOpt}}=0
\end{equation}
which denotes that the battery resistance should be chosen as close
to zero as possible.

\subsection{Step 11: Battery Weight Optimization}

\subsubsection{Optimal Battery Nominal Voltage $U_{\text{bOpt}}$}

As analyzed in \textit{Section\,\ref{subsec:motorWeight}}, the actual
working voltage of the motor is determined by the battery voltage,
and the constraint in Eq.\,(\ref{eq:UbTmax}) should be satisfied
to make sure the motor has the minimum weight. Therefore, after the
optimal motor NMV $U_{\text{mMaxOpt}}$ is determined, the optimal
battery voltage $U_{\text{bOpt}}$ is also determined as 
\begin{equation}
U_{\text{bOpt}}=U_{\text{mMaxOpt}}.\label{eq:UbOpt}
\end{equation}

\subsubsection{Optimal Battery Capacity $C_{\text{bOpt}}$ }

After the above procedures, the propeller, motor, ESC and battery
of the propulsion system both have the maximum efficiency, which means
the battery has the minimum current $I_{\text{b0}}$ under the hovering
mode with propeller thrust $T_{\text{hover}}$. According to \cite{Shi2017},
with knowing the parameters $\Theta_{\text{pOpt}}^{*},\Theta_{\text{mOpt}}^{*},\Theta_{\text{eOpt}}^{*},U_{\text{bOpt}},T_{\text{hover}},n_{\text{p}},h_{\text{hover}}$,
the battery current $I_{\text{b0}}$ can be estimated through the
propulsion system equivalent circuit in Fig.\,\ref{Fig02}. By substituting
the desired propeller thrust $T_{\text{hover}}$ into the propeller,
motor, ESC and battery models in Eqs.\,(\ref{Eq02PropTorque})(\ref{Eq05Um})(\ref{Eq06Ie})(\ref{Eq08Ue})
successively, the battery discharge current $I_{\text{b0}}$ can be
obtained. Then, the optimal battery capacity $C_{\text{bOpt}}$ can
be obtained by substituting $t_{\text{discharge}}=t_{\text{hover}}$
into Eq.\,(\ref{eq:tdischarge}), which yields that 
\begin{equation}
C_{\text{bOpt}}=\frac{t_{\text{hover}}I_{\text{b0}}}{0.85}\cdot\frac{1000}{60}\label{EqCbOpt}
\end{equation}

\subsubsection{Optimal \textmd{MDR $K_{\text{bOpt}}$}}

The battery MDR $K_{\text{b}}$ should be chosen as small as possible
within the constraint in Eq.\,(\ref{eq:KbConstraint}). Therefore,
by substituting the obtained $U_{\text{mMaxOpt}}$ and $C_{\text{bOpt}}$
into Eq.\,(\ref{eq:KbConstraint}), the optimal battery MDR K$_{\text{bOpt}}$
is given by 
\begin{equation}
K_{\text{b}}=\frac{1000\left(n_{\text{p}}I_{\text{{mMaxOpt}}}+I_{\text{other}}\right)}{C_{\text{bOpt}}}.\label{eq:KBop}
\end{equation}

\subsection{Step 12: Optimal Battery Product}

With the obtained parameters $\Theta_{\text{bOpt}}=\left\{ U_{\text{bOpt}},K_{\text{bOpt}},C_{\text{bOpt}},R_{\text{bOpt}}\right\} $,
the optimal battery product can be determined by searching the battery
product database. The parameter set of the obtained optimal battery
product is represented by $\Theta_{\text{bOpt}}^{*}=\left\{ U_{\text{bOpt}}^{*},K_{\text{bOpt}}^{*},C_{\text{bOpt}}^{*},R_{\text{bOpt}}^{*}\right\} $.

Similar to the selection principles in \textit{Section\,\ref{SecOptMotSel}},
the optimal battery product should be determined from the battery
database by comparing the parameters in the sequence $U_{\text{bOpt}}^{*},K_{\text{bOpt}}^{*},C_{\text{bOpt}}^{*},R_{\text{bOpt}}^{*}$,
and each parameter in $\Theta_{\text{bOpt}}^{\ast}$ should be chosen
equal or close to the corresponding parameter in $\Theta_{\text{bOpt}}$
with satisfying the constraints 
\begin{equation}
\begin{array}{l}
U_{\text{bOpt}}^{\ast}=U_{\text{bOpt}},\,K_{\text{bOpt}}^{\ast}\geq K_{\text{bOpt}}\end{array}.
\end{equation}
Noteworthy, the battery voltage $U_{\text{bOpt}}^{\ast}$ should satisfy
the constraint that $U_{\text{bOpt}}^{\ast}=U_{\text{bOpt}}=U_{\text{mMaxOpt}}$
to ensure that the motor can work under the desired voltage.

In practice, designers have to build a battery pack to satisfy the
above design requirements by connecting small battery cells in series
or parallel. According to \cite[pp. 46]{quan2017introduction}, by
combining battery cells in series, a higher voltage can be obtained,
with capacity unchanged. On the other hand, by combining battery cells
in parallel, larger capacity and discharge current can be obtained,
with voltage unchanged.

\section{Experiments and Verification}

\label{Sec-6}

\subsection{Statistical Model Verification}

\label{Sec-6-1}
\begin{figure}[tbh]
\centering \includegraphics[width=0.45\textwidth]{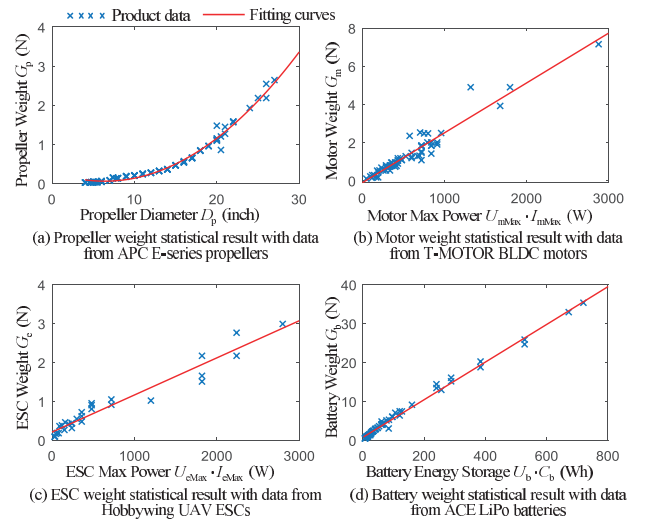} \caption{Weight statistical results for the propulsion system components.}
\label{Fig09-0} 
\end{figure}

Comprehensive statistical analyses for the products
of propellers, motors, ESCs and batteries on the market are performed
to verify the weight statistical functions $f_{G_{\text{p}}}\left(\cdot\right)$,
$f_{G_{\text{m}}}\left(\cdot\right)$, $f_{G_{\text{e}}}\left(\cdot\right)$
and $f_{G_{\text{b}}}\left(\cdot\right)$ in Eqs.\,(\ref{eq:Gp})(\ref{eq:GmOrgin})(\ref{eq:GeOrgin})(\ref{eq:Gb})
respectively. Some typical results are presented in Fig.\,\ref{Fig09-0},
where the products come from four most well-known manufacturers (APC,
T-MOTOR, Hobbywing, Gens ACE). Fig.\,\ref{Fig09-0} shows the relationship
between the weight and the parameters of each component. The statistical
results are consistent with the analysis results in \textit{Section\,}\ref{Sec-4}.

\subsection{Optimization Method Verification}

\label{SecPropDiam-1}

According to Eq.\,(\ref{Eq15DHOpt0}), the optimal pitch angle for
T-MOTOR series propellers (parameters are listed in Eq.\,(\ref{Eq04KCM}))
is obtained as $\varphi_{\text{pOpt}}=0.1054$. For the comparative
validation, a series of multicopter propellers from the website \cite{TMotor2017}
are listed in Table\,\ref{Table2TProp}. It can be observed from
Table\,\ref{Table2TProp} that the statistical diameter/pitch ratio
result is $D_{\text{p}}/H_{\text{p}}\approx3$ ($\varphi_{\text{p}}\approx0.1065$),
which is in good agreement with the theoretical result $\varphi_{\text{pOpt}}=0.1054$.
The comparison between the calculation value and the statistical value
shows that the pitch angle optimization method is effective to find
the optimal pitch angle adopted by manufacturers. Meanwhile, the propeller
test data from the UIUC website \cite{UIUC2017} also show that the
obtained pitch angle $\varphi_{\text{pOpt}}$ can guarantee high efficient
in increasing the thrust and decreasing the torque.

\begin{table}[th]
\caption{Diameter/pitch ratio of carbon fiber propellers from the website \cite{TMotor2017}.}
\label{Table2TProp} \centering%
\begin{tabular}{|c|c|c|c|c|}
\hline 
Propeller  & P12x4  & P14x4.8  & P15x5  & P16x5.4\tabularnewline
\hline 
$D_{\text{p}}/H_{\text{p}}$  & 3  & 2.92  & 3  & 2.963\tabularnewline
\hline 
\hline 
Propeller  & P18x6.1  & $\cdots$  & G26x8.5  & G28x9.2\tabularnewline
\hline 
$D_{\text{p}}/H_{\text{p}}$  & 2.951  & $\cdots$  & 3.06  & 3.04\tabularnewline
\hline 
\hline 
Propeller  & G30x10  & G32x11  & G34x11.5  & G40x13.1\tabularnewline
\hline 
$D_{\text{p}}/H_{\text{p}}$  & 3  & 2.91  & 2.96  & 3.05\tabularnewline
\hline 
\end{tabular}
\end{table}

The motor T-MOTOR U11 KV90 is adopted as an example to verify the
proposed optimization method. The calibrated motor parameters of  U11
KV90 are listed below

\begin{equation}
\begin{array}{c}
K_{\text{V}}=90\text{RPM/V},\text{ }U_{\text{mMax}}=48\text{V},\text{ }I_{\text{mMax}}=36\text{A},\\
U_{\text{m0}}=10\text{V},\text{ }I_{\text{m0}}=0.7\text{A},\text{ }R_{\text{m}}=0.3\Omega
\end{array}\label{eq:U11Param-1}
\end{equation}
and the corresponding experiment results from \cite{TMotor2017} are
listed in Table\,\ref{Tab4PropData}.

\begin{table}[ptb]
\caption{Full-throttle test data of U11 KV90 with 12S Li-Po battery (48V)}
\label{Tab4PropData}\centering%
\begin{tabular}{|c|>{\centering}p{0.04\textwidth}|>{\centering}p{0.05\textwidth}|>{\centering}p{0.035\textwidth}|c|>{\centering}p{0.05\textwidth}|>{\centering}p{0.04\textwidth}|}
\hline 
Prop.  & Current (A)  & Power (W)  & Thrust (N) & RPM  & Torque (N$\cdot$m)  & Tempe. ($^{\circ}$C) \tabularnewline
\hline 
27x8.8CF  & 24.6  & 1180.8  & 81.4  & 3782  & 2.623  & 58.5\tabularnewline
\hline 
28x9.2CF  & 28.3  & 1358.4  & 91.3  & 3696  & 3.068  & 66.5\tabularnewline
\hline 
29x9.5CF  & 31.9  & 1531.2  & 98.8  & 3602  & 3.41  & 78.5\tabularnewline
\hline 
30x10.5CF  & 36.3  & 1742.4  & 106.8  & 3503  & 3.846  & HOT$!$\tabularnewline
\hline 
\end{tabular}
\end{table}

From the perspective of theoretical calculation, the optimal diameter
of motor U11 KV90 can be obtained by substituting the motor parameters
in Eq.\,(\ref{eq:U11Param-1}) into Eq.\,(\ref{Eq19DpOpt}), where
the obtained result is $D_{\text{pOpt}}=29.7$\,inches. Therefore,
according to the constraint in Eq.\,(\ref{eq:DpOpt}), the propeller
diameter should be chosen as ${D_{\text{pOpt}}^{\ast}}=29$\,inches
because the motor will overheat if the propeller diameter is larger
than $D_{\text{pOpt}}$.

It can be observed from the experimental results in the last two rows
of Table\,\ref{Tab4PropData} that the motor temperature becomes
overheated when the propeller diameter changes from 29\ inches to
30\ inches. Therefore, the optimal propeller diameter obtained from
experiments should be 29\,inches, which agrees with the theoretical
optimal solution ${D_{\text{pOpt}}^{\ast}}=29$\,inches.

\subsection{Design Optimization Example}

As an example, assume that the given design task is to select an optimal
propulsion system for a multicopter ($n_{\text{p}}=4$) whose total
weight $G_{\text{total}}=196\text{\,N}$ (20\ kg) and flight time
is $t_{\text{hover}}=17$\,min (flight altitude $h_{\text{hover}}=50\,\text{m}$).
The component databases are composed of the ESC, BLDC motor, caber
fiber propeller products from T-MOTOR website \cite{TMotor2017},
and the LiPo battery products from GENS ACE website \cite{ACE2017}.

The key calculation results of the design procedures are listed as
follows.

i) The thrust requirements for the propulsion system are obtained
from Eqs.\,(\ref{eq:T0})(\ref{Eq31T0TpmaxThr}) as $T_{\text{hover}}=49$\,N
and $T_{\text{max}}=98$\,N, where $\gamma=0.5$ is adopted here.

ii) The optimal propeller efficiency parameters are obtained from
Eqs.\,(\ref{Eq14BpOpt})(\ref{Eq15DHOpt0}) as $B_{\text{pOpt}}=2$,
$\varphi_{\text{pOpt}}=0.1065$\,rad. Then, with statistical models
in Fig.\,\ref{Fig05ThustWattVolt-1} and Fig.\,\ref{Fig05ThustWattVolt},
the optimal motor parameters are obtained from Eqs.\,(\ref{Eq36UmMax})(\ref{Eq37ImMaX1})(\ref{Eq37KVOpt})
as $U_{\text{mMaxOpt}}=48$\,V, ${{I}_{\text{mMaxOpt}}}=34$\,A,
${{K}_{\text{VOpt}}}=91$\,RPM/V. Therefore, by searching products
from the T-MOTOR website according to principles in Eqs.\,(\ref{eq:secPrin})(\ref{EqMotorSeq}),
the optimal motor is determined as U11 KV90.

iii) With the blade parameters in Eq.\,(\ref{Eq04KCM}), the optimal
propeller diameter can be obtained from Eq.\,(\ref{Eq19DpOpt}) as
$D_{\text{pOpt}}=0.7468\text{\,m}$, and the optimal propeller product
is selected as 29x9.5CF 2-blade. In the same way, the ESC parameters
can also be obtained from Eq.\,(\ref{eq:OptESC}) as $U_{\text{eMaxOpt}}=48$\,V
and ${{I}_{\text{eMaxOpt}}}=34$\,A, and the optimal ESC product
is selected as FLAME 60A HV.

iv) For the battery, the optimal parameters are obtained from Eqs.\,(\ref{eq:UbOpt})(\ref{EqCbOpt})(\ref{eq:KBop})
as $U_{\text{bOpt}}=48$\,V, $C_{\text{bOpt}}=16000$\,mAh and $K_{\text{bOpt}}=10$\,C.
The optimal battery selected from ACE website is TATTU LiPo 6S 15C
16000mAh $\times$ $2$.

The obtained result has been verified by several multicopter designers.
Experiments show that a quadcopter with the designed propulsion system
can efficiently meet the desired design requirements. If a larger
motor (like T-MOTOR U13) is selected, then a smaller propeller has
to be chosen for generating the same full-throttle thrust $T_{\text{pMax}}$,
which reduces the motor efficiency because the optimal operating condition
cannot be reached. As a result, the obtained propulsion system is
heavier than the optimal result according to our experiments. If a
smaller motor is selected, then a larger propeller has to be chosen,
which results in exceeding the safety current of the motor. Therefore,
the obtained propulsion system is optimal within the given database.

\subsection{Method Application}

If the same optimization problem is solved by brute force searching
methods to traverse all combinations and evaluate the performance
of each combination, the time consumption will be far longer than
the proposed method. For example, it takes about 100ms for our evaluation
method in \cite{Shi2017} to calculate the performance of each propulsion
combination. Assuming that the numbers of products in the propeller,
motor, ESC and battery databases all equal to $n$, then it will take
about $T\left(n\right)\approx C_{n}^{4}=O(n^{4})$ to traverse all
products in the databases. By comparison, the computation amount of
the proposed method to traverse 4 component databases is 4$n$. Since
there are 15 parameters in Table\,\ref{tab1}, it is easy to verify
that the total computation amount of the proposed method is approximate
to $T\left(n\right)\approx15\cdot4n=O(n)$, which is much faster than
the brute force searching methods.

The optimization algorithm proposed in this paper is adopted as a
sub-function in our online toolbox (URL: \uline{\url{www.flyeval.com/recalc.html}})
to estimate the optimal propulsion system with giving the multicopter
total weight. The program is fast enough to be finished within 30\ ms
by using a web server with low configuration (single-core CPU and
1GB of RAM). The feedback results from the users show that the optimization
results are effective and practical for the multicopter design.

\section{Conclusions}

\label{Sec-7}

In this paper, the precise modeling methods for the propeller, ESC,
motor and battery are studied respectively to solve the optimization
problem for the propulsion system of multicopters. Then, the key parameters
of each component are estimated through mathematical derivations to
make sure that the obtained propulsion system has the maximum efficiency.
Experiments and feedback of the website demonstrate the effectiveness
of the proposed method. The propulsion system is the most important
part of a multicopter, and its design optimization method will be
conducive to the fast, optimal and automatic design of the whole multicopter
system or other types of aircraft systems. The theoretical analysis
can be further used to directly maximize the endurance of all kinds
of UAVs, which is interesting for future research.

\bibliographystyle{IEEEtran}
\bibliography{IEEEabrv}

\begin{biographynophoto} {Xunhua Dai} received the B.S. and M.S.
degrees from Beihang University, Beijing, China, in 2013, and 2016,
respectively. Currently, he is a Ph.D. candidate of School of Automation
Science and Electrical Engineering at Beihang University, Beijing,
China. His main research interests are reliable flight control, model-based
design and design optimization of UAVs.

\end{biographynophoto}

\begin{biographynophoto} {Quan Quan} received the B.S. and Ph.D.
degrees from Beihang University, Beijing, China, in 2004, and 2010,
respectively. He has been an Associate Professor with Beihang University
since 2013, where he is currently with the School of Automation Science
and Electrical Engineering. His research interest covers reliable
flight control, vision-based navigation, repetitive learning control,
and time-delay systems. \end{biographynophoto}

\begin{biographynophoto} {Jinrui Ren} received the B.S. degree
from Northwestern Polytechnical University, Xi\textquoteright an,
China, in 2014. Currently, She is a Ph.D. candidate of School of Automation
Science and Electrical Engineering at Beihang University, Beijing,
China. Her main research interests include nonlinear control, flight
control, and aerial refueling. \end{biographynophoto}

\begin{biographynophoto} {Kai-Yuan Cai} received the B.S., M.S.,
and Ph.D. degrees from Beihang University, Beijing, China, in 1984,
1987, and 1991, respectively. He has been a full professor at Beihang
University since 1995. He is a Cheung Kong Scholar (chair professor),
jointly appointed by the Ministry of Education of China and the Li
Ka Shing Foundation of Hong Kong in 1999. His main research interests
include software testing, software reliability, reliable flight control,
and software cybernetics. \end{biographynophoto}
\end{document}